# Geo-neutrinos: recent developments

Stephen T. Dye [a]

[a] *Hawaii Pacific University, 45-045 Kamehameha Hwy, Kaneohe 96744, U.S.A.*

**Abstract**

Radiogenic heating is a key component of the energy balance and thermal evolution of the Earth. It contributes to mantle convection, plate tectonics, volcanoes, and mountain building. Geo-neutrino observations estimate the present radiogenic power of our planet. This estimate depends on the quantity and distribution of heat-producing elements in various Earth reservoirs. Of particular geological importance is radiogenic heating in the mantle. This quantity informs the origin and thermal evolution of our planet. Here we present: currently reported geo-neutrino observations; estimates of the mantle geo-neutrino signal, mantle radiogenic heating, and mantle cooling; a comparison of chemical Earth model predictions of the mantle geo-neutrino signal and mantle radiogenic heating; a brief discussion of radiogenic heating in the core, including calculations of geo-neutrino signals per pW/kg; and finally a discussion of observational strategy.

## 1. Introduction

Geo-neutrinos are produced during the nuclear beta decay of isotopes in the decay chains of uranium and thorium, which are the main heat-producing elements in the Earth. The heat generated by these decays contributes to mantle convection and plate tectonics, resulting in earthquakes and volcanoes. Measuring the amount and location the uranium and thorium provides fundamental information on the formation and thermal evolution of our planet. Geo-neutrino observations provide a unique method for directly estimating radiogenic heating in the mantle. The level of internal heating in this inaccessible Earth reservoir has significant implications for models of the origin and thermal evolution of our planet.

## 2. Current geo-neutrino observations

Underground observatories, one in Japan and one in Italy, are currently measuring the rate of geo-neutrino interactions. These observatories do not discern the interactions of uranium geo-neutrinos from those of thorium geo-neutrinos. Measurements of the geo-neutrino signal rate assume the ratio of thorium atoms to uranium atoms in the Earth is 3.9. They are typically given in terrestrial neutrino units (TNU). A TNU is one detected electron antineutrino interaction per free proton target every $10^{32}$ years. At Japan the measured signal is 30±7 TNU, using an exposure of 4.90 $TNU^{-1}$ collected since 2002 [1]. At Italy the measured signal is 39±12 TNU, using an exposure of 0.37 $TNU^{-1}$ collected since 2007 [2].

The reported observations identify the Earth as the source by recording the energy spectrum, not the direction, of interacting antineutrinos. The energy spectrum is found to be consistent with geo-neutrinos from the decay of uranium and thorium. There is ambiguity with how much of the observed signal originates from each Earth reservoir. Geochemistry demands negligible amounts of uranium and thorium in the core [3]. This requirement, which is discussed below in the context of the geo-dynamo, predicts an insignificant geo-neutrino signal from the core. The typical analysis assumes the crust and mantle are the only sources of geo-neutrinos. An extraction of the mantle signal then becomes a simple subtraction of the predicted crust signal from the total signal.

## 3. Observed mantle radiogenic heating

The reported geo-neutrino observations estimate the radiogenic heating of the mantle. This estimate requires a prediction of the crust signal. A study of the crust near the observatory in Japan predicts 26 TNU [4]. Applying an uncertainty of 15% gives a crust signal of 26±4 TNU. Subtracting the crust prediction from the total observed at Japan gives a mantle signal of 4±8 TNU, assuming uncorrelated errors. A detailed study of the crust near the observatory in Italy predicts 25±4 TNU [5]. Subtracting the crust prediction from the total observed at Italy gives a mantle signal of 14±13 TNU, assuming uncorrelated errors. A weighted average of these measurements finds a mantle signal of 7±7 TNU. The observatories in Japan and Italy are beginning to resolve the mantle geo-neutrino signal.

Radiogenic heating in the mantle, corresponding to a geo-neutrino signal of 7±7 TNU, depends on the distribution of uranium and thorium in the mantle. Mantle structure with radial symmetry customarily employs a preliminary reference earth model (PREM) [6]. Using the seismically determined mantle structure of PREM and assuming a homogeneous distribution of heat-producing element concentrations estimates 8±8 TW of mantle radiogenic heating, following established methods [7]. This estimate assumes constant heat-producing element ratios throughout the mantle (Th/U=3.9; K/U=12,000). Inhomogeneous distributions of heat-producing element concentrations producing the same geo-neutrino signal can yield different levels of radiogenic heating. For example, placing material enriched in heat-producing elements in a layer coincident with the seismic structure D″ at the base of overlying mantle depleted in heat-producing elements can produce 7 TNU of geo-neutrino signal and 10 TW of radiogenic power. The possibility of different levels of radiogenic heating arising from plausible distributions of uranium and thorium in the mantle introduces systematic uncertainty to the estimate of radiogenic heating in the mantle based on geo-neutrino observations.

The uncertainty in the present estimate of mantle radiogenic heating is dominated by the errors on the total measured geo-neutrino signal rates not on the subtracted crust signal predictions. Improved precision of the total signal rates requires increased observational exposure.

## 4. Rate of mantle cooling

Heat leaves the planet at a measured rate of 47±3 TW [8]. This total comprises heat leaving the core at a rate of 13±3 TW [9], the lithosphere at a rate of 8±1 TW [10], and the mantle. The mantle component includes radiogenic heating, which is estimated by geo-neutrino observations at a minimum rate of 8±8 TW, and mantle cooling. Simple subtraction, assuming uncorrelated errors, estimates mantle cooling at a rate of 18±9 TW. The probability that mantle cooling is not present is less than a few per cent. Uncertainty in the estimate of mantle cooling is dominated by the relatively poor precision of mantle radiogenic heating from geo-neutrino observations. This precision, which depends on the uncertainty of the total signal, is expected to improve with increased exposure to geo-neutrinos.

## 5. Predictions of chemical Earth models

Chemical models of the Earth provide predictions of mantle radiogenic heating. Constraints on mantle radiogenic heating from geo-neutrino observations could potentially favour one model over another, providing clues to the origin of the Earth. A recently revised model of the Earth based on the chemical composition of enstatite chondrites predicts 6.5±1.6 TW of mantle radiogenic heating and a corresponding mantle geo-neutrino signal rate of 5.1±1.0 TNU [11]. An often-referenced model of the Earth based on the chemical composition of carbonaceous chondrites predicts 12.3±4.5 TW of mantle radiogenic heating and a corresponding mantle geo-neutrino signal of 9.1±3.0 TNU [12,13,10], according to the calculations of this author. The predictions of both these models are consistent with present observations of geo-neutrinos and are marginally consistent with each other, considering uncertainties.

## 6. Radiogenic heating in the core

Radiogenic heating in the core is not unfamiliar to models of the geo-dynamo powered only by thermal energy [14]. Recent evaluations of the thermal and electrical conductivity of iron at core pressure and temperature suggest that heat flows out of the core at a rate of ~13 TW [15,16]. Some models find acceptable solutions by incorporating radiogenic heating at levels of several pW/kg [15]. The concentration, mass, geo-neutrino flux, geo-neutrino signal, and power for each

Table 1

Calculated quantities for the main heat-producing isotopes for each pW/kg of internal heating in the outer core are listed. $a$ is concentration, $m$ is mass, $\Phi$ is flux, $S$ is geo-neutrino signal, and $P$ is power.

|  | $a$(X) ng/g | $m$(X) $10^{16}$ kg | $\Phi$(X) $10^5$ cm$^{-2}$s$^{-1}$ | $S$(X) TNU | $P$(X) TW |
|---|---|---|---|---|---|
| $^{238}$U | 10.1 | 1.87 | 1.63 | 2.14 | 1.84 |
| $^{232}$Th | 38.1 | 7.00 | 1.29 | 0.52 | 1.84 |
| $^{40}$K | 35.1 | 6.47 | 17.1 | n.a. | 1.84 |

of the heat-producing isotopes per pW/kg are listed in Table 1.

Present detection methods do not resolve the source of geo-neutrinos, making it currently not possible to distinguish core signal from mantle signal. Separation of these signals could become possible with the development of sensitivity to geo-neutrino direction. In this case, resolution on the order of the half angle subtended by the core, approximately 30 degrees, would be required. The possibility of radiogenic heating in the core due to uranium or thorium introduces additional systematic uncertainty to the estimate of radiogenic heating in the mantle based on geo-neutrino observations.

## 7. Discussion

The geo-neutrino signal from the mantle, which relates directly to the radiogenic heating of this inaccessible reservoir, is a key to understanding fundamental geological processes. An unresolved distribution of uranium and thorium within the mantle introduces uncertainty associated with geo-neutrino estimates of mantle radiogenic heating at the few TW level. Observations from continental sites carry an additional uncertainty due to the crust signal interfering with the mantle signal. An estimate of this uncertainty, taken from above, is ±4 TNU, corresponding roughly to ±4 TW. Moreover, radiogenic heating of the outer core by uranium at a level of a pW/kg produces about 2 TNU of geo-neutrino signal, introducing another potential source of uncertainty in the estimate of mantle radiogenic heating. Taking the quadratic sum of these errors suggests systematic uncertainty of ±5 TW for the geo-neutrino estimate from a single continental site.

Avoiding the continental interference with the mantle signal suggests the observation of geo-neutrinos from the great oceans [17]. Systematic uncertainty associated with the distribution of uranium and thorium within the mantle and core remains. This contributes about ±3 TW of uncertainty to the estimate of mantle radiogenic heating based on an observation from an oceanic site. In addition to deployment risks, a deep-ocean observation entails unaddressed technical challenges, such as light output and attenuation of scintillating liquid at low temperature and high pressure.

An alternative for reducing continental interference with the mantle signal is to expand the existing network of underground observatories. Indeed, a third geo-neutrino observatory is under preparation in eastern Canada [18]. Taking the weighted average of mantle geo-neutrino measurements reduces the crust signal systematic uncertainty, potentially by a factor of the inverse square root of the number of independent measurements. The optimal strategy for observing mantle geo-neutrinos depends on the precision of mantle radiogenic heating required to advance geological knowledge.

## Acknowledgments

This work was supported in part by National Science Foundation grants through the Cooperative Studies of Earth's Deep Interior program (EAR 0855838 and EAR 1068097) and by various awards from Hawaii Pacific University to promote faculty scholarship.

## References

[1] A. Gando et al., Phys. Rev. D88 (2013) 033001.
[2] G. Bellini et al., Phys. Lett. B722 (2013) 295.
[3] W.F. McDonough in: R.W. Carlson (Ed.), Treatise of Geochemistry, vol. 2, Elsevier, Oxford, 2003, pp. 547-568.
[4] S. Enomoto et al., Earth Planet. Sci. Lett. 258 (2007) 147.
[5] M. Coltorti et al., Geochim. Cosmochim. Acta 75 (2011) 2271.
[6] A.M. Dziewonski, D.L. Anderson, Phys. Earth Planet. Inter. 25 (1981) 297.
[7] S.T. Dye, Rev. Geophys. 50 (2012) RG3007.
[8] J.H. Davies, D.R. Davies, Solid Earth 1 (2010) 5.
[9] H. Gomi et al., Phys. Earth Planet. Inter. 224 (2013) 88.
[10] Y. Huang et al., Geochem. Geophys. Geosyst. 14 (2013) 2013.
[11] M. Javoy, E. Kaminski, Earth Planet. Sci. Lett. 407 (2014) 1.
[12] W.F. McDonough and S. s-. Sun, Chem. Geol. 120 (1995) 223.
[13] R. Arevalo Jr. et al., Earth Planet. Sci. Lett. 278 (2009) 361.
[14] T. Lay et al., Nature Geosci. 1 (2008) 25.
[15] M. Pozzo et al., Nature 485 (2012) 355.
[16] H. Gomi et al., Phys. Earth Planet. Inter. 224 (2013) 88.
[17] C.G. Rothschild et al., Geophys. Res.Lett. 25 (1998) 10830.
[18] M. Chen, Earth Moon Planets 99 (2006) 221.